# A STATE OF THE ART ON RECENT PROGRESS AND EMERGING CHALLENGES ON ENERGY TRANSFER BETWEEN VIBRATING MODES UNDER AN EXTERNAL MECHANICAL FORCE WITH TIME-VARYING FREQUENCY FROM 2020 TO 2025

José Manoel Balthazar [1, 2]*; Jorge Luis Palacios Felix [3], Mauricio A. Ribeiro[1], Angelo Marcelo Tusset [1],Jeferson José de Lima[1], Vinicius Piccirillo[1], Julijana Simonović [4,] Nikola D. Nešić [5], Marcos Varanis [6], Clivaldo de Oliveira[7] , Raphaela C. Machado[8], Gabriella O M Oliveira [1]

[1] Federal University of Technology—Parana, Ponta Grossa, Paraná, Brazil
[2] Faculty of Mechanical Engineering, São Paulo State University, Bauru, SP Brazil
[3] Federal University of the Southern Border - UFFS, Cerro Largo campus
[4] Faculty University of Nis, Nis, Serbia
[5] Faculty of Technical Sciences, University of Priština in Kosovka Mitrovica, Serbia
[6] Federal University of Mato Grosso do Sul, Campo Grande/MS, Brazil
[7] Federal University of Grande Dourados, Dourados/MS, Brazil.
[8] Faculty of Electrical Engineering, São Paulo State University *Guaratinguetá, SP, Brazil*

*Correspondent author José Manoel Balthazar <jmbaltha@gmail.com>*

## Abstract

*In this paper, we will discuss an example of nowadays importance with a future perspective on engineering, which excitations sources have always limited power, limited inertia and their frequencies vary according to the instantaneous state of the vibrating system. The practical examples of non-ideal systems are considered. The most common phenomenon for this kind of system is discussed. The period consider is from 2020 to 2025. The specific properties for various models are also discussed. The directions of the future investigation are given. In this paper, the authors revisited some publications based on the assumptions that the external excitations are produced by Nonideal sources (RNIS), that is of limited power supply. Among these applications nonlinear phenomena such as the Sommerfeld effect, Saturation phenomenon was observed, considering fractional damping, Energy Harvester and Jacobi-Anger expansion was used in the governing equations of motion. We also used Jacobi-Anger expansion in the case of energy transfer between vibrating modes under external force with time varying frequency that will the future directions of the research on nonideal vibrating systems (RNIS).*

1. Introduction

The features commonly found in vibrating engineering are nonlinearities (geometric or physical characteristics); Dissipation of energy (internal or external); Gyroscopic Systems; Imperfections of material; Stationary and non-stationary modes; Unlimited power sources (or Ideal sources) or limited (non-ideal sources (RNIS)) and non-conservative dynamic (follower) or forced. It is also well known that Ideal systems (IS) are vibrating systems that consider the actuator influence over the system on which he is working, but do not consider the influence of the structure to the actuator. Conversely, systems that consider the dynamic influence of the structure on the actuator receive the name of non-ideal systems (RNIS). The first problem characterized as non-ideal (RNIS) to be studied was the Sommerfeld effect, in 1904. When analyzing an electromechanical system, consisting of a beam and an electric motor, Arnold Sommerfeld realized that the system had some instability, reacting differently when passing the resonance regions. It was noticed that when the system came closer to the resonance region, the energy supplied to the system was not fully converted to angular speed of the motor, but was lost, serving only to increase the amplitude of the structure vibration. After passing through the resonance region, the amplitude of vibration falls, and the motor angular frequency rises with the increase in motor voltage. This effect is also known as jump phenomenon (Konenenko, 1969).

The development of modern technology, machinery and equipment is becoming more and more complex every day. Among all machinery, we will restrict ourselves to a particular rotary one and the full interaction of their support structures. Rotary machinery and the full interaction with their support structures are a major, and critical component, of many mechatronic systems, in industrial plants, aerial and ground transportation vehicles, and in many other applications on modern engineering and applied sciences. The rotor imbalance is the most common reason in rotary machinery to present undesirable vibrations. Most of the rotary machinery problems may be solved by using the rotor balancing and misalignment. The mass imbalance in a rotating system often produces excessive synchronous forces, which reduces the life span of various mechanical elements. A very small amount of imbalance may cause severe problems in high-speed rotary machinery. The vibrations caused by unbalance may destroy critical parts of the machine, such as bearings, seals, gears, and couplings. Rotor unbalance is a condition, in which the center of mass of a rotary assembly, typically the shaft and its fixed components, like disks and blades etc., does not coincident with the center of rotation. In practice, rotors can never be perfectly balanced because of manufacturing errors such as porosity in casting, non-uniform density of material, manufacturing tolerances and gain or loss of material during operation. As a direct result of mass imbalance, a centrifugal force is generated and must be reacted against by bearing and support structures. In this way, some phenomena were observed in a composed vibrating system supporting structures and rotating machines, where the disbalance of the rotating parts is the major cause of the vibrations.

It is well known that in the design of structures; it is necessary to investigate the relevant dynamics to predict the structural response due to the excitations. The integration of mechanical, electromagnetic, and computer elements (electro-mechanical) to produce devices and systems that monitor, and control machines and structural systems has led to the need for integration of mechanical and electrical design, and industrial applications to nonlinear mechatronic systems **(Industry 4.00).**

This state-of-the-art survey aims to provide a short forum for the discussion and dissemination of the latest approaches, methodologies results and current challenges in nonlinear vibrations, and in the field of electro-mechanical systems (EMS).(EMS) is not just a marriage of electrical and mechanical systems and is more than just a control system; it is a complete integration of all of them. Topics of interest on (EMS) include Interdisciplinary approaches and complex nonlinear phenomena in problems encountered in emergent engineering and science practice. Therefore, this paper attempts to provide a short review of some of the latest efforts in the development and applications of recent exciting research progress of rotary non-ideal systems (RNIS) that is considering the full interaction between (RNIS) with their support structures.

Here, this main purpose is to provide the researchers and engineers, who are working in this emergent vibrating field, with comprehensive knowledge to help them with better applying of the theory of rotary non-ideal systems (RNIS), to solve problems related to rotary machines and the interaction with the support structure. and nowadays trends show a marked tendency for solutions of vibrating systems analysis of drive systems as early as the design stage. So, this work is an overview of the current (and nowadays) literature dealing with the main properties of non-ideal vibrating systems (RNIS). The most common phenomenon for these kinds of systems is discussed, that is, the Sommerfeld effect and saturation phenomenon.

We announced that new and old studies of (RNIS), with limited power supply (small DC motors or electrodynamical shakers), are usually used in laboratory tests, and therefore, the investigation of mutual interactions of driven and driving sub-systems is very important. In this paper, the main properties of (RNIS) have been reviewed, such as the Sommerfeld effect, i.e., jump phenomena and the increase in power supply that is required by an excitation source operating near resonance. The possibility of saturation phenomenon occurrence, i.e., the transference of energy from higher frequency and lower amplitude to lower frequency and higher amplitude mode; and the existence of regular (periodic motion) and irregular (chaotic or hyperchaotic motion) behaviors, depending on the value of control parameters (voltage of a DC motor).In this paper, we will show that: It is of major importance to consider non-ideal energy sources (RNIS) in engineering problems (mechanical, electrical, aerospace and so on): They acted on an oscillating system and at the same time experienced reciprocal action from the system. In the DC motor-frame system with slow increase of power levels the Sommerfeld's effect appears in resonance, as there is not enough power to reach higher speed regimes with lower energy consumption). Saturation of high frequency low amplitude mode and transference of energy to low frequency high amplitude mode is possible if the non-ideal system frequency is twice the natural frequency of frame i.e. 2:1 internal resonance. We also can highlight the influence of the linear and nonlinear coupling parameter of the piezoelectric material has great influence in collecting energy harvester in the non-ideal. Chaos control is possible to be done. The control may be directed at the oscillator and not at the motor by controlling the voltage of the DC motor as is usually done.

In this point, it is important to clarify what rotary non-ideal systems (RNIS) means, to avoid future confusions.  Non-ideal systems (RNIS), have appeared in the literature with several meanings; as an example: some researchers use the concept of (RNIS) solutions for concentrated solutions, that is, the solutions can occur in two ways: when intermolecular forces between solute and solvent molecules are less strong than between molecules of similar (of the same type) molecules, and when intermolecular forces between dissimilar molecules are greater than those between similar molecules. Here, we deal with energy transfer between the energy sources and the support structures and their possible control approaches, that is, we are interested in what happens to the motor (or electro-mechanical shaker), input, output, as the response to the rotary system support structure changes. So, in this paper, practical examples of non-ideal systems are considered. The most common phenomenon for this kind of system is presented. The specific properties for various models are also discussed. The directions of the future investigation are given. Several references based on the analysis of resonant behavior of non-ideal vibrating systems were published in the current literature. The appearance of chaos in (RNIS) was also reported by several authors in current literature. In this paper, the period considered is from **_2020 to 2025_**, because before, many publications were posted on this subject before.

The reader may see some examples of comprehensive details and different approaches to (RNIS) vibrating problems through some publications, entirely devoted to the subject. We mention the classical books by (Kononenko,1969),(Evan-Iwanowski,1976),(Alifov,Frolov,1990), (Blekhman,2000), and recent by ( Cveticanin *et al.* 2017) , (Balthazar *et al.*,2022, 2025).Some relevant papers published by (Dimentberg, *et al.*, 1997), ( Balthazar *et al.,* 2003), ( Balthazar et al.,2011),( Felix *et al.*, 2016) ( Balthazar *et al.,* 2016 ), ( Balthazar *et a.l*, 2018) , ( Balthazar det al, 2022), underserved of others .

We will organize this paper as follows: In section 2, we discuss classical modeling of (RNIS). In section 3, a brief literature review of classical modeling of (RNIS) is discussed. In section 4, we present nowadays formulation of (RNIS). In section 5, we mention the concluding remarks in this paper. Finally, we present the acknowledgements, and the main bibliographic references used.

## 2. Classical modeling of (RNIS)

The (classic) modeling and governing equation of motion of (RNIS) are given as follows.

If we consider a classical vibrating system driven by a harmonic excitation $f= f_0 \cos \omega t$ , having frequency $\omega$ and amplitude $f_0$ (in this case, they are constants), the so-called dynamics of the ideal system (IS), may be described by the following classical differential equation:

$$\frac{d^2x}{dt^2} + 2\varsigma \frac{dx}{dt} + \omega_0^2 x = f_0 \cos \omega t \qquad (1)$$

where $\omega_0^2$ is the natural frequency and $\varsigma$ is damping coefficient of the vibrating system.

On the other hand, for non-ideal dynamical systems (RNIS), one must add an equation that describes how the energy source supplies the energy to the equations, which governs the corresponding ideal dynamical system (IS).

In Figure 1, we consider total mass total mass $M = m_0 + m_1$ . The $c\vartheta$ is the angular viscous damping coefficient and J is the rotor inertia moment $m_0$ is the unbalanced mass and r is the eccentricity. Elastic support stiffness is k and cx is the viscous damping coefficient. Rotor angular displacement is $\vartheta$ and motor linear displacement is x.

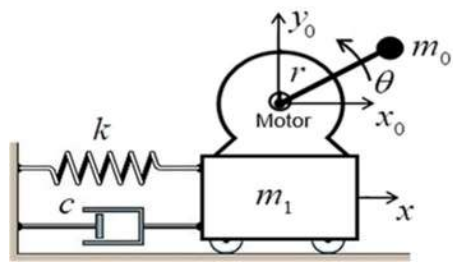

Figure 1. (RNIS) motor with unbalanced rotor and its elastic support.

So, the Lagrange's equation for $x$ and $\theta$ coordinates the governing equations of motion are:

$$M\ddot{x} + c\dot{x} + kx - m_0 r \ddot{\theta} \sin\theta - m_0 r \dot{\theta}^2 \cos\theta = 0 \quad (2)$$

$$(J + m_0 r^2)\ddot{\theta} + c_\theta \dot{\theta} - m_0 r \ddot{x} \sin\theta + mgr \cos\theta = T_e - T_L \quad (3)$$

where $T_e$ is the motor torque and $T_L$ is the load torque. To write the governing equations of motion equations, considering dimensionless parameters, one defines:

$$\zeta_x = \frac{c_x}{M}, \omega_0 = \sqrt{\frac{k}{M}}, \mu = \frac{m_0 r}{M}, \zeta_\theta = \frac{c_\theta}{(J+m_0 r^2)}, \varepsilon = \frac{m_0 r}{(J+m_0 r^2)}, \alpha = \frac{1}{(J+m_0 r^2)} \quad (4)$$

And considering the nonlinear damping $F_d = sgn(\dot{x})|\dot{x}|^n$ with n=Fractional order and $T_e - T_L = T(\dot{\theta})$ and $\Omega_0$ = angular velocity of the motor, $a$ = constant of motor torque and $b$ = coefficient angular da straight line, $f_0 = \frac{\Omega_0}{2.\pi}$ [Hz] ( Figure 2)

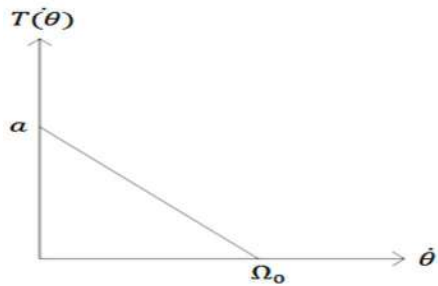

Figure 2: Torque of the motor

we will obtain:

$$\ddot{x} = \frac{-c_x}{M} \cdot (Sign\ \dot{x} \cdot ([\dot{x}]^n\ ) - \frac{k}{M}x + \frac{m_0 r}{M}(\ddot{\theta}\sin(\theta) + \dot{\theta}^2 \cos(\theta))$$

$$\ddot{\theta} = \frac{a}{(m_0 r^2 + J)} - \frac{b}{(m_0 r^2 + J)} \cdot \dot{\theta} + \frac{m_0 r}{(m_0 r^2 + J)} \cdot (\ddot{x}.\operatorname{sen}\theta - g.\cos\theta) \qquad (5)$$

Figures 3 a, b exhibits the amplitude of vibration versus $f_0$ (Sommerfeld effect) for various values of n, considering initial conditions nulls and the parameter given by Table 1.

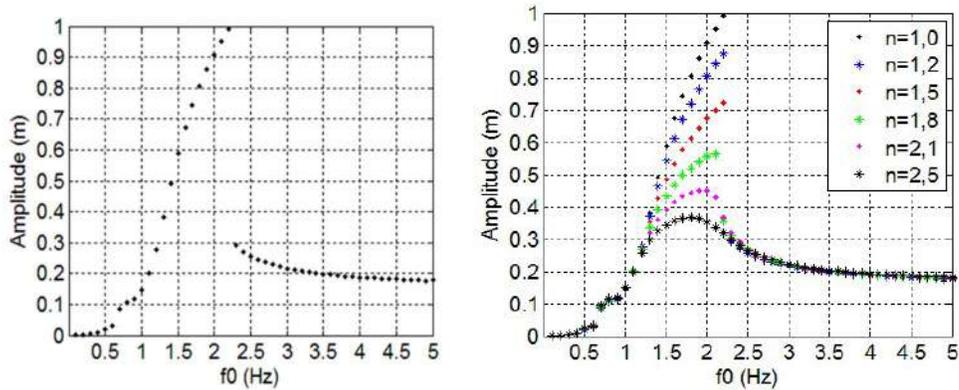

Figures 3 : amplitude of vibration versus $f_0$ (Sommerfeld effect):  Fig 3a, n=1. Fig 3b various values of n.

Table 1. Values of the parameters used in Figure 3

| Parameter | Value |
| --- | --- |
| M [kg] | 5 |
| m.e [kg.m] | 1 |
| k [N/m] | 533 |
| c [N.s/m] | 5 |
| a [N.m] | 80 |
| J [kg.m²] | 0,1 |
| g [m/s²] | 9,81 |

Using governing eq (1), in the next topic (3), we will discuss a brief literature review to nowadays approaches to (RNIS) published from **2020 to 2025**.

3. **Literature review of classical modeling (RNIS) from 2020 to 2025**

Using the classical formulation given by section 2, we present next, a brief literature review of some relevant papers, throughout the period from 2020 to 2025 , undeserved of others.

The authors (Lebedenko and Mikhlin, 2025) dealt with the resonant behavior of two 3-DOF systems with a limited power supply (or so-called non-ideal systems). Resonant steady states are constructed by using the multiple scales method. The obtained analytical results are compared with a numerical simulation with success. It is also shown that amplitudes of the resonance vibrations of the elastic sub-system can be essentially decreased by change of system parameters.

A Non-Linear Control and Numerical Analysis Applied in a Non-Linear Model of Cutting Process subject to a non-ideal excitation was studied by (Tusset *et al.,* 2024). The authors considered that cutting tool vibration in the machining process is one of the main problems of productivity and machining accuracy. Introduction of a magnetorheological damper was also considered in the proposed model to reduce the vibration amplitudes of the cutting tool and suppress the chaotic behavior. Hysteresis was also considered in the magnetorheological damper (MR) model and its application in the system as both a passive and active absorber. The active control strategy considered the application of two non-linear control signals: feedforward to maintain the vibration with a desired behavior and state feedback to drive the system to the desired behavior. The numerical results demonstrated that the proposed controls efficiently reduced the vibration amplitude by introducing the MR damper. Active control has proven effective in controlling the force of the MR damper by varying the electrical voltage applied to the damper coil.

The use of limited power motor fixed to flexible structures is widely used in engineering and has seen a significant increase in published studies. These components form a non-ideal system, which can cause undesirable phenomena throughout its execution, such as the capture of parametric oscillation by resonance, studied by Sommerfeld, whose effect bears his name. One study, on a vibration absorber in a non-ideal system, was done in (Petrocino et al. 2023), on a model that represents a fixed beam and at the free end a motor with its unbalanced shaft. In this work, based on the equations that govern the movement, an analytical development was carried out, using the average method, to obtain an approximate solution and numerical simulations with the absorber on and off. The results demonstrated a reduction in parametric excitation, an effect caused by the electromagnetic. On the other hand, in( Petrocinio *et al*., 2024, the results obtained before, of the tests were compared with the results of the models from the revisited work, validating the performance of the passive electromagnetic vibration absorber

The study done by (González-Carbajal et al., 2023) addressed the influence of the Sommerfeld effect on the behaviour of vibrocompacting machines driven by unbalanced electric motors. Based on a simplified model with 4 degrees of freedom , the authors, they investigated the dynamics of quasistatic vibro compaction processes, i.e. those in which the compaction was achieved through a sufficiently slow increase in the input power of the electric motors. The main novel contribution of the paper was the characterization of the evolution of the system state through a quasistatic torque curve, which was analytically derived.

In (Yaroshevich et al, 2023) the dynamics of an unbalanced rotor with a vibrating suspension axis and driven by an asynchronyous electric motor of limited power were considered. Stationary (near stationary) modes of rotation of the rotor with a frequency equal to the vibration frequency of the axis were investigated. An explanation of the phenomenon of vibrational capture of rotation of an unbalanced rotor was given. The proposed mechanical interpretation of the effect allowed deeper understanding of the classical results and conclusions. The obtained condition for the existence of a stationary mode allowed us to estimate the frequency capture interval of the rotor. The case when the mode of vibration capture of rotation is not set is considered. For such a case, an expression for the vibrational moment is obtained, as well as an equation for slow motions. Attention was drawn to the possibility of occurrence in the considered modes of motion of slow (relative to the rotation frequency) rotor oscillations with sufficiently large amplitudes. It was demonstrated that the vibrational capture mode has the property of self-regulation; allowing the rotation frequency of an unbalanced rotor during load oscillations. Attention was drawn to the fact that in this mode of motion, there is certainly a transfer of energy either from the source of vibration to the rotor, or vice versa. The Sommerfeld effect in an oscillatory system with an inertial vibration exciter is represented by vibration capture of rotation of the vibration exciter by resonant oscillations of the carrier body. The theoretical results were confirmed by numerical modelling.

In the paper (Pechuk *et al.*, 2023), the authors addressed energy transmission via an electro-dynamical transducer from the amplifier and energy harvesting from wave field. In the first case an amplifier was considered as a self-exciting system with a limited power. Electrical current produced by it was converted by the transducer into mechanical force, which leads to vibrations of the base. A mechanical oscillator is mounted on the transducer base. The influence of oscillator vibrations on the formation of the driving force leads to the Sommerfeld effect. Expressions for supplied and consumed powers were shown. The energy harvesting problem was also discussed. The classical results for wave power harvesting by wave energy extractor as a single degree of freedom system were presented in the second considered problem. The example included an axisymmetric buoy which oscillates and was subjected to its natural hydrostatic restoring force. Main attention was focused on the values and expressions for the mean powers. The expression for the maximum mean power was given for the system considered.

Nonlinear modal analysis is a powerful technique that allows a better understanding of the nonlinear dynamic behavior of mathematical models with few degrees of freedom. Nonlinear modes have emerged as a natural extension of linear modes for systems with large oscillation amplitudes. (Piccirillo et al, 2022) applied the technique of nonlinear normal modes analysis in a nonlinear Duffing oscillator driven by a limited (non-ideal) power source. Results showed that the use of nonlinear modal analysis allowed the description of the complex behavior of the system, resulting from the nonlinearities considered.

In (Djanan, 2022) the physical parameters of the (RNIS) structure enabled to present the phase and anti-phase or rapid and late synchronization phenomena between the motors. This difference of phase or the input delay between the motors and the voltage applied on the motors leads to situations where the amplitude vibrations of the mechanical structure are considerably reduced.

( Yaroshevich et al, 2022) taking into account that stationary (near stationary) rotation modes of an unbalanced rotor with the oscillation frequency of its axis were considered. It was demonstrated that the vibration moment arising as a result of oscillations tends to synchronize vibrational and rotational movements. In addition, its effect in the stationary mode becomes unambiguous – either braking or accelerating. It is shown that when a stationary rotation mode is established, large, relatively slow oscillations of rotor's speed can occur. Theoretical results are confirmed by computer simulation.

(González-Carbajal et al, 2022) investigate the dynamics of a Duffing oscillator excited by an unbalanced motor. The interaction between motor and vibrating system was considered as non-ideal(RNIS), which means that the excitation provided by the motor can be influenced by the vibrating response, as is the case in general for real systems. This constitutes an important difference with respect to the classical (ideally excited) Duffing oscillator, where the amplitude and frequency of the external forcing are assumed to be known as a priori. Starting from pre-resonant initial conditions, we investigate the phenomena of *passage through resonance* (the system evolves towards a post-resonant state after some transient near-resonant oscillations) and *resonant capture* (the system gets locked into a near-resonant stationary oscillation). The stability of stationary solutions is analytically studied in detail through averaging procedures, and the results obtained are confirmed by numerical simulations.

The authors in (Brogin et al, 2022)proposed a new approach to suppressing Sommerfeld effect these undesired effects, based on fuzzy Takagi-Sugeno Modeling, in a three-degree-of-freedom shear building considering the equations of motion with and without a non-linear stiffness. The results show that the vibrations can be efficiently controlled, and even chaotic patterns of motion were suppressed.

In (Djanan et al, 2022) the authors presented the effects of late switching on (time delay) between two or three DC electrical machines characterized by limited power supplies on their fast or late self-synchronization when mounted on a rectangular plate with simply supported edges. The DC electrical machines were considered here as non-ideal oscillators, rotating in the same direction and acting as an external excitation on a specific surface of the plate. The stability analysis of the whole studied system (with two machines) around the obtained fixed point was done through analytical and numerical approaches by using the generalized Lyapunov and Routh-Hurwitz criterion.

The article ( Iskakov *et a.l*, 2022) examined the effect of linear damping and combined linear and nonlinear cubic damping of an elastic support on the dynamics of a gyroscopic rigid rotor with a non-ideal energy source, considering cubic nonlinear stiffness of the support material. Analysis of the research results showed that both linear damping and combined linear and nonlinear cubic damping can significantly suppressed the resonance peak of the fundamental harmonic, reduced the amplitude of vibration frequency variation and stabilized the shaft rotation speed, but the effect of combined damping is more significant. In non-resonant regions, where the speed is higher than the natural frequency of the rotor system, both types of damping shorten the distance between jumps in nonlinear resonance curves and eliminate them.

 (Varanis, *et al.*, 2021) in this article, concisely written to be an introducing tutorial comparing different time-frequency techniques for non-stationary signals. The theory was carefully exposed and complemented with sample applications on mechanical vibrations and nonlinear dynamics.

A particular phenomenon that was also observed in non-stationary systems is the Sommerfeld effect, which occurs due to the interaction between a non-ideal energy source and a mechanical system. An application through TFA for the characterization of the Sommerfeld effect was presented

(Pechuk et al, 2021) considered a new non-ideal (RNIS)modified cardiorespiratory model based on the famous DeBoer beat-to beat model and Zaslavsky map (which describes dynamics of the respiratory system as a generator of central type) was studied in detail. The respiratory tract was first modeled by a self-oscillating system under the impulsive influence of heartbeat and cardiovascular system was represented as an oscillating system with a limited excitation. Chaotic modes were revealed, which were produced by the interaction between the subsystems. It was proved that the irregularity of the behavior of phase trajectories depends on the intensity of the effect produced by the heart rate on breathing, which is characteristic of the dynamics of the cardiovascular system of a healthy person.

In ( Mikhlin,et al, 2020) Three-DOF system with a limited power supply (or non-ideal system) having the Mises girder as absorber was considered. Stationary resonance regimes of vibrations near stable equilibrium positions of the system were constructed in two approximations of the multiple scales method. Namely, vibrations near the resonance 1:1 between the motor rotation frequency and the linear sub-system frequency and vibrations near the resonance 1:1 between the motor and absorber frequencies are analyzed. This analysis near the first resonance was made as in supposition that elastic springs of the Mises girder are weak and have an order of the small parameter, as well as in supposition that these springs are not weak. It was obtained that the effective absorption of the elastic vibrations can be obtained in the second case. Additional numerical simulation permits to find regimes of absorption when it is possible to guarantee also the fast outcome from the first resonance region. The most appropriate for absorption and such outcome from the first resonance were e vibrations near the second resonance and the snap-through motion

( Varanis *et al.*, 2020) dealt with a non-ideal system with memory by possessing a fractional damping term. The system is characterized by additional cubic nonlinearity. To distinguish between periodic and non-periodic behaviors, they are used three different mathematical tools, which are the 0–1 test, scale index and wavelet technique.

(Kong, *et al*., 2020 in their paper, considered the Sommerfeld effect and self-synchronization of two non-ideal induction motors in a simply supported beam system are studied. Based on fully considering non-ideal characteristics of induction motors and interactions between the beam structure and induction motors, a continuous model was developed and discretized by using the assumed mode method.

In the paper done by ( Rocha etal, 2020),the authors consider the application of the piezoelectric energy collection using a portal structure of two-degrees-of-freedom. The piezoelectric material was considered as a linear device using a capacitive mathematical model. The portal structure is of two-degrees-of-freedom, considering a quadratic coupling between the first and second modes of vibration. 2:1 internal resonance between the first and second modes is set, which is a special condition of this type of system due to the appearance of a saturation phenomenon. The results show considerably nonlinear behavior due to the non-ideal motor and, with the saturation phenomenon, it is more efficient to collect energy by coupling the PZT to the column. The stability of the system due to the piezoelectric coefficient T is also accounted for analysis, as it is related to the coupling of the structure. The stability analyzes are carried out by numerical tools as phase planes, Poincare maps bifurcation diagrams, 0-1 test.

( Sinha *et al*., 2020) in their study on non-ideal problems used an analytical prediction for jumps obtained from a steady-state power balance whereas the transient analysis was performed with the help of Bond Graph (BG) models and MSC-ADAMS software. The transient responses from the numerical models were used to verify the analytical results with success.

(Bharti, *et al.,* 2020) considered the Sommerfeld effect of the first kind relates to the resonance at the synchronous rotor whirl (critical speeds), whereas that of the second kind relates to instability of the rotor whirl. Due to such nonlinear jumps, some of the speed ranges where the rotor is theoretically stable under ideal or mathematical conditions may not be reached in practice, i.e. with a real drive which is naturally non-ideal. The dynamics of this rotor system are analytically and numerically studied in this article. The numerical simulations are performed with a multi-energy domain bond graph model which guarantees energetic consistency of the model. Finally, in (Piccirillo et al, 2020), considered the dynamic integrity analysis on a non-ideal oscillator. The dynamic integrity of the periodic solution was s studied and the basin erosion is evaluated. The erosion profiles obtained allow us to identify the practical thresholds that guarantee a priori and a safe project to be developed.

4. **Nowadays modeling formulation of (RNIS), using Jacobi-Anger expansion,**

In this section we will deal with frequency-varying excitations (RNIS) Equations**, inspired** on the paper ( Felix et al, 1976). The basic idea is s outline, next.

Using an averaging process:

$$x = a(t)\cos(\theta(t) + \beta(t)), \quad \dot{\theta} = \Omega(t) \qquad (6)$$

we will obtain the expressions of modulation of amplitude and phase, and the variable angular velocity can be approximated by (Kononenko 1969):

$$\frac{d\theta}{dt} = \Omega_0 + a_0 \cos(b_0 \Omega_0 t + c_0) \qquad (7)$$

where $\Omega_0$ is a constant obtained from averaging of the angular frequency at resonance.

Note that the parameters $a_0$, $b_0$ and $c_0$ are defined by the active interaction between the oscillating system and the excitation source. Furthermore, they are considered as control parameters.

If $a_0 = 0$, then $q_3 = \varphi = \Omega_0 t$ corresponding to the classical harmonic excitation.

Then, the Eq (3) may be written as only one equation:

$$\ddot{x} + 2\varsigma_1 \cdot (Sign\,\dot{x} \cdot ([\dot{x}]^n) + \omega_1^2 x = f_0 \cos[\Omega_0 t + a_0 \sin(b_0 \Omega_0 t + c_0)] \qquad (8)$$

Where the constant $f_0$ is the amplitude of excitation in horizontal direction respectively with excitation frequency $\Omega_0$..

Using the classical Jacobi-Anger expansion, the sine and term can be written as:

$$cos(z \sin\theta) = \sum_{k=-\infty}^{\infty} J_k(z) \cos(k\theta) \tag{9}$$

where $J_k(z)$ is the k-th Bessel function. The Bessel functions (Fig 4) on the first kind and positive integer $k$ are defined as:

$$J_k(x) = \sum_{n=0}^{\infty} \frac{(-1)^n}{n!\,(n+k)!} \left(\frac{x}{2}\right)^{2n+k}$$

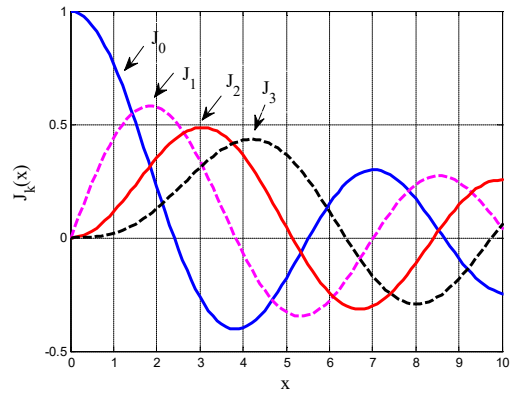

**Figure 4. First four Bessel Functions**

So, using the trigonometric function can de written as:

$$cos[\Omega_0 t + a_0 \sin(b_0 \Omega_0 t + c_0)] = \sum_{k=-\infty}^{\infty} J_k(a_0) \cos(\Omega_k t + k c_0)$$

With $\Omega_k = \Omega_0 + k b_0 \Omega_0$, and we will obtain (Felix et al, 2016):

$$\ddot{x} + 2\varsigma_1 2\varsigma_1 . (\,Sign\, \dot{x}\cdot([\dot{x}\cdot]^n\,) + \omega_1^2 x = f_0 \sum_{k=-\infty}^{\infty} J_k(a_0)\cos(\Omega_k t + kc_0) =$$

$$f_0 J_k(a_0)\cos(\Omega_k t + kc_0) \qquad (15)$$

It is important to note that the Eq. (15) is like a system with harmonic excitation for each $k$. It is a generalization of earlier results in the current literature.

On the other hand, if the system, Fig 5 , has two degrees of freedom such as a modeling of a nonlinear portal frame, where the excitation is an eccentric rotating mass forced by a small DC motor mounted rigidly on the top. The flexible portal frame structure having two columns clamped at their bases, with height $h$ and constant cross section of area $A_c$ and moment of inertia $I_c$, with concentrated weights at the tops of the columns of mass $m$. The horizontal beam is pinned to the columns at both ends with length $L$ and constant cross section, of area $A_b$ and moment of inertia $I_b$. A linear elastic material is considered whose young modulus is $E$. The vibration modes of the portal frame in generalized coordinates. $x = q_1$ is related to the horizontal displacement of the central section of the beam in the sway mode (with natural frequency $\omega_1$) and $y = q_2$ to the mid-span vertical displacement of the same section of the beam in the first symmetrical mode (with natural frequency $\omega_2$).

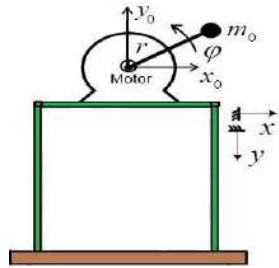

**Figure 5. Schematic of vibration modes and eccentric mass motion**

In the schematic representation of ( Fig 5), the excitation was an eccentric mass $m_0$ at a distance $r$ from the axis drive by small DC motor placed at the mid-span of the beam. The angular displacement of this mass is adopted as generalized coordinate

$$\varphi = \Omega_0 t + a_0 \sin(b_0 \Omega_0 t + c_0) \quad (7a)$$

The control of passing through resonances or resonance capture of the angular velocity ($\frac{d\varphi}{dt} \approx \omega_2$) is driven by a rotational actuator $u(t)$ with various constant torques levels. is the coefficient of resisting torque. The governing equations of motion characterizing the portal frame:

$$\ddot{x} + 2\varsigma(sign(\dot{x})|\dot{x}|^n) + \omega_1^2 x + \alpha_1 xy = f_0 \cos[\Omega_0 t + a_0 \sin(b_0 \Omega_0 t + c_0)]$$

$$\ddot{y} + 2\varsigma(sign(\dot{y})|\dot{y}|^n) + \omega_2^2 y + \alpha_2 x^2 = g_0 \sin[\Omega_0 t + a_0 \sin(b_0 \Omega_0 t + c_0)]$$

(16a)

Or using the classical Jacobi-Anger expansion

$$\ddot{x} + 2\varsigma(sign(\dot{x})|\dot{x}|^n) + \omega_1^2 x + \alpha_1 xy = f_0 \sum_{k=-\infty}^{\infty} J_k(a_0) \cos(\Omega_k t + kc_0)$$

$$= g_0 J_k(a_0) \cos(\Omega_k t + kc_0)$$

$$\ddot{y} + 2\varsigma(sign(\dot{y})|\dot{y}|^n) + \omega_2^2 y + \alpha_2 x^2 = g_0 \sum_{k=-\infty}^{\infty} J_k(a_0) \sin(\Omega_k t + kc_0)$$

$$= g_0 J_k(a_0) \sin(\Omega_k t + kc_0)$$

(16b)

Again, this is like a vibrating system with harmonic excitation for each $k$.

### 4.1 Extension of Governing equations of motion to (RNIS) excited by an electrodynamic shaker

Another form of (RNIS) governing equation of motion may be, obtained, when one considers the action of an electro-dynamical shaker (excited the support structure of (RNIS) which, consists of a device with the objective to reproduce a harmonic excitation by considering a mechanical and an electrical part. In this case, the interaction of a support structure of (RNIS) and the electrodynamic shaker, is done by the existence of quasi-periodic oscillations (Sommerfeld effect) ( Balthazar et al, 2018).

We will need to consider the quantity $T$ (transducer , constant), which relates the current in the coil to the magnetic force on the considered coil. The transducer constant is given by $T = 2\pi nlB$, where $n$ is the number of turns in the coil, $l$ is the radius of the coil, and $B$ is the uniform radial magnetic field strength in the annular gap. The transducer $T$ also relates the electrical potential $e$, across to the terminals of the coil to the velocity of the coil, with respect to the permanent magnet. The expression of the voltage, over the resistor and the capacitor may be written respectively, as a nonlinear function of the instantaneous electrical charge q:

$$V_R = -Ri_0 \left( \frac{\dot{q}}{i_0} - \left( \frac{\dot{q}}{i_0} \right)^3 \right) \qquad (17)$$

$$V_C = \frac{q}{C} + aq^3 \qquad (18)$$

where $i_0$ is an initial current, in the electrical part, and $a$ is the nonlinear coefficient, that depends on the type of the capacitor, in use.

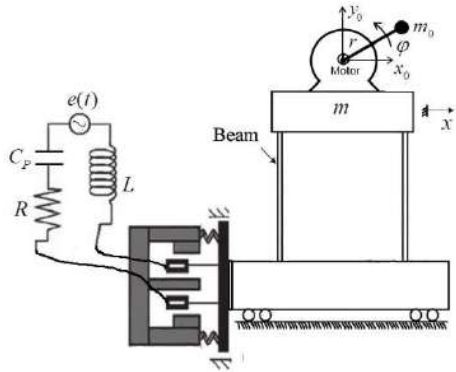

Figure 6- Schematic of the support (RNIS) excited by an electrodynamic shaker

In this case, the governing equation of motion are:

$$\ddot{x} + f(x,\dot{x}) - T\dot{q} = f_0 \cos(\Omega_0 t + a_0 \sin(b_0 \Omega_0 t + c_0))$$

$$L\ddot{q} - R\left(1 - \frac{1}{i_0^2}\dot{q}^2\right)\dot{q} + \frac{1}{C}q + aq^3 + T\dot{x} = 0 \quad (19)$$

Where $x = (x_1, x_2)$

The extension to another devices is easy to do.

We announced that several papers were published using this new formulation. We mention some of them, next.

**4.1 A Brief State of the art of Nowadays formulation of (RNIS)using Jacobi-Anger expansion**

Here, the state-of-the-art refers to the latest ideas and methods of nowadays formulation of (RNIS).

(Silva *et al.,* 2025) examined the nonlinear dynamic behavior of the coupling interactions between three oscillators that describe the regions of the human heart responsible for electrical impulses. Specifically, we focused on the sinoatrial node, the atrioventricular node, and the Purkinje complex. The dynamics of cardiac behavior can be effectively represented by a nonlinear dynamic model, such as the Van der Pol Oscillator. However, we added an external force described by Bessel functions applied to oscillators with non-ideal characteristics to explore the system's behavior further.

In the paper ( Ribeiro et al, 2024) the authors explored a non-linear dynamic mathematical model containing Shape Memory Alloy (SMA) and a non-ideal engine for energy production. The authors found chaotic behavior for a given set of parameters. Then, in this paper was applied two control techniques applied to suppress the chaotic behavior for a desired periodic orbit. The first is the State-Dependent Riccati Equation (SDRE) which considered the non-linearities of the system and Optimal Linear Feedback Control (OLFC) which employed a linear methodology to control the device. The results were promising due to the trajectory errors between the controllers that show that chaos was suppressed, and the current produced by the system became periodic.

In (Ribeiro et al, 2024), the authors dealt with cargo transport to the space station and the transport of satellites to orbit the Earth and the use of rockets with propellant fuel has been growing. In this paper the authors proposed a mathematical model to analyze nonlinear dynamic behavior. Such a model is based on non-ideal Mathieu's solution. However, we consider that the application of force for rocket propulsion is not ideal. An obtained result, the parametric set was established to diagnose the chaotic and periodic behavior that can be observed in mathematical modeling. Establishing these analyzes allows the application of future work in the development of control projects to reduce vibrations that can cause anomalies in the trajectories for the system's entry into orbit.

( Ribeiro *et al.* , 2023) , explored the nonlinear dynamics for capturing energy from a device, which is considered support with cantilever beam ferromagnetic containing piezoceramic patches connected to an electrical circuit that collects energy for capture. The free end of the ferromagnetic cantilever beam is under the effect of two magnetic poles, and they consider an asymmetric potential for the magnetic poles. However, the influence of asymmetry in bi-stable energy collectors can change the potential well and adjust the distribution of its potential energy. Charging in the potential well also modifies the unstable equilibrium positions thus altering the dynamic characteristics of the device and thus benefiting the use of energy under various excitation conditions. Furthermore, asymmetric potentials are combined with the movement of human lower limbs for applications in energy harvesting devices in such displacement. As numerical results, we explored the nonlinear dynamics and established the sets of parameters for the convergence of the trajectories and the regions of maximum average output power.

The paper ( Ribeiro et al, 2023) presented a new approach to analyzing the initial conditions of an energy harvesting system under the action of a non-ideal motor that was described in terms of its angular displacement. The analyses were based on the motor rotation frequency. However, for our analysis, we used the permutation entropy that determined the regions in which the initial conditions are more complex for the system for some values of the motor frequency. This result corroborated those found by the maximum Lyapunov exponent and the output power. We also analyzed the dynamics of two initial conditions that meet in two regions, one with high entropy values and the other with low entropy values. Therefore, we observe the nonlinear dynamics behavior (Bifurcation diagrams and maximum Lyapunov exponent). With these analyses, we proposed an orbit with higher energy and application of an optimal linear feedback control (OLFC) design to stabilize the system in this orbit and optimize energy harvesting.

In the work of (Ribeiro et al.,2022) , the authors analyzed the nonlinear fractional dynamics in the equations of motion of a bar coupled to support under the effect of a potential described by two equally spaced magnetic poles. We also considered Bouc–Wen damping in the equations of motion. For external force vibrations, we considered an equation of a non-ideal motor based on the parameters that related the interaction between the oscillation and the excitation source. With such considerations, we explored the influence of the fractional derivative operator parameter on the average power generated by the device and the dynamic behavior to determine the chaotic and periodic regions. As a conclusion, we established a set of parameters for the fractional differential equations to obtain higher average powers and the periodicity windows that corroborate the establishment of energetic orbits for energy harvesting.

Finally, we mention that ( in Nesli et al., 2025)the authors using numerical methods, specifically the Incremental Harmonic Balance (IHB) and used Jacobi-Anger expansion in the investigations the vibration of the platform structure subjected to excitation of non-ideal motor, with success.

## 5. concluding remarks

In this paper, we analyzed the dynamical coupling between energy sources ( of limited power supply) and structural response, which must not be ignored in real engineering problems, since the majority real motors have limited output power. We also used Jacobi-Anger expansion in the case of energy transfer between vibrating modes under external force with time varying frequency that will the future directions of the research on nonideal vibrating systems(RNIS).

## Acknowledgements

All of the authors acknowledges CNPq: Conselho Nacional de Desenvolvimento Científico e Tecnológico, Brazilian foundation.

**credit authorship contribution statement:** All authors also contributed to the execution of this work

**Conflicts of Interest:** The authors declare no conflicts of interest.

## Basic References